\documentclass[sigconf,nonacm]{acmart}

\AtBeginDocument{%
  }

\usepackage{multirow}
\usepackage{subcaption}
\usepackage{makecell}


\acmISBN{xxx}

\begin{document}

\title{Demystify, Use, Reflect, Assess (DURA): An Experience Report on LLM Integration in CS2}

\author{Margaret Ellis}
\affiliation{%
  \institution{Virginia Tech}
  \city{Blacksburg}
  \country{US}}
\email{maellis1@vt.edu}
 
\author{Nikitha Donekal Chandrashekar}
\affiliation{%
  \institution{Virginia Tech}
  \city{Blacksburg}
  \country{USA}}
\email{nikitha@vt.edu}

\author{Sehrish Basir Nizamani}
\affiliation{%
  \institution{Virginia Tech}
  \city{Blacksburg}
  \country{USA}}
\email{sehrishbasir@vt.edu}

\author{Mohammed Farghally}
\affiliation{%
  \institution{Virginia Tech}
  \city{Blacksburg}
  \country{USA}}
\email{mseddik@vt.edu}
 
\author{Jake O'Brien}
\affiliation{%
  \institution{Virginia Tech}
  \city{Blacksburg}
  \country{USA}}
\email{jakeobrien@vt.edu}

\author{Naren Ramakrishnan}
\affiliation{%
  \institution{Virginia Tech}
  \city{Alexandria}
  \country{USA}}
\email{naren@vt.edu}

 \balance
\begin{abstract}

Student access to Large Language Models (LLMs) is reshaping learning behaviors; at the same time students are entering the workforce where effective LLM use is becoming an expected skill.
In this Experience Report we share our DURA framework (Demystify-Use-Reflect-Assess) and materials we used to restructure our CS2 course to allow the use of LLMs. We first demystified LLMs, then provided guidance on use with required attribution. We also added reflections related to LLM use at three points throughout the semester to encourage student meta-cognition around LLM use. We increased the value of proctored assessments in tandem with allowing retakes and including questions that explicitly assess skills from programming assignments. Students reported using LLMs for clarifying course concepts, debugging, understanding assignment guidelines, and determining test cases, but also still sought assistance via office hours and TAs, monitored Piazza, and reviewing course content. 
Students articulated thoughtful and strategic approaches to LLM use and also valued the instructional content and guidance from course staff. Student use of office hours increased slightly this semester and student perceptions that the instructor cares about them and their learning improved.

\end{abstract}


\maketitle

\section{Introduction and Motivation}
In May of 2025, we saw the rise of AI coding agents for software development and tech company layoffs. The Computing Community Consortium (CCC) and the Computing Research Associations- Industry (CRA-I) white paper "The Future of Programming in the Age of Large Language Models" recommended educators move towards emphasizing skills needed to work collaboratively with LLMs such as decomposing problems and testing code for correctness {~\cite{GuhaZorn2025FutureProgramming}}.

\begin{figure}[ht]
  \centering

  \includegraphics[width= 0.85\linewidth]{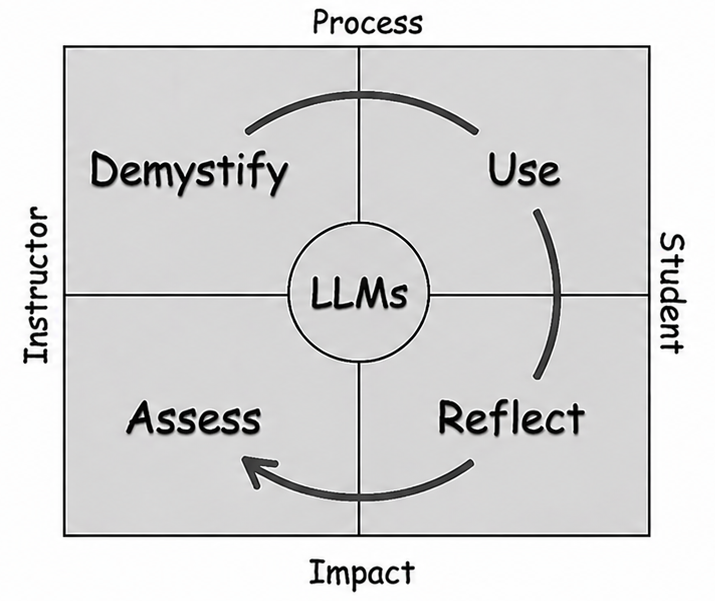}
  \caption{DURA (Demystify-Use-Reflect-Assess), a lightweight framework for updating courses to incorporate AI/LLMs}

  \Description{Demystify
	Instructor on the left and Student on the right with a circle through four quadrants from Demystify to Use to Reflect to Assess. Process across the top labelling Demystify and Use and Evidence across the bottom labeling Assess and Reflect.

.}
  \label{fig:framework}
\end{figure}

In response, we considered how to update our CS2 course, which already emphasized decomposing problems and unit testing, to also prepare our students to use LLMs and still be able to meet the course learning outcomes without LLM assistance. We considered updates according to DURA (Demystify-Use-Reflect-Assess) as shown in Table \ref{tab:dura_cs2}, our lightweight framework for incorporating LLMs into courses. 
The four components of DURA: demystify, use, reflect, and assess, deliberately support the learning process, as well as the need for reliable evidence of student learning. 
The foundational component of DURA is the demystification of Large Language Models (LLMs), which underpin modern AI. When students understand how the technology works, they are more likely to be effective AI users who won’t overly trust the models or attribute thoughts and feelings to them!  Students can also benefit from instruction and practice to use LLMs responsibly and productively. Beyond use, students are expected to reflect on their experiences with LLMs, supporting metacognition and self-regulation. Given the rapidly changing landscape and increasing student access to LLMs, instructors must also consider authentic approaches for assessment that align with these realities. 
The DURA strategies we used are highlighted in Table \ref{tab:dura_cs2} and some or all of which can be applied to most CS courses.

\begin{table}[t]
\caption{DURA framework implementation in CS2}
\label{tab:dura_cs2}
\centering
\small
\begin{tabular}{|l|p{0.72\linewidth}|}
\hline
\textbf{Component} & \textbf{Our CS2 Implementation} \\
\hline
\textbf{Demystify} & Demystify how LLMs work, how they can affect learning and performance. \\
\hline
\textbf{Use} & Use LLMs with guidance and attribution. \\
\hline
\textbf{Reflect} & Reflect on LLM use and problem-solving process. \\
\hline
\textbf{Assess} & Assess programming assignment skills through proctored assessments and allow retakes. \\
\hline
\end{tabular}
\end{table}

Prior work suggests the availability of LLMs may negatively affect course climate and reduce student use of course forums and office hours \cite{amoozadeh2024student,bernstein2025beyond}. We had observed such a downward trend in our CS2 course over the past few years and there was slight uptick during the semester when these changes were implemented. We found that students' perceptions that the instructor cared about them and their learning increased during the semester when we implemented these updates. 

The goal of this experience report is to describe how a course can be redesigned to incorporate student use of LLMs while preserving core learning outcomes and promoting effective learning behaviors.  We aim to 

\begin{enumerate}
    \item Describe our implementation of the DURA (Demystify-Use-Reflect) framework in a large CS2 course,
    \item Examine how students used LLMs, particularly in relation to help-seeking and problem-solving,
    \item Examine how these course changes may be associated with student perceptions of the course.
\end{enumerate}
Our intention is to provide practical guidance and reusable materials for instructors adapting their CS courses to incorporate LLMs.

\section{Prior and Related Work}
\subsection{Demystify}
 In recent years, many AI-literacy frameworks for K-12 have been published. For example, Jia et al. developed a research based framework for developing a holistic AI literacy framework for children based on three dimensions: AI awareness, AI mechanics, and AI impacts \cite{Jia2025ailiteracy}. Mills et al. previously published a detailed framework for K-12 to understand, evaluate, and use AI. These frameworks both emphasize an understanding of how AI works before using AI. The Digital Education Council provides a five dimension framework for AI literacy in higher education, each at three competency levels: 1) Understanding AI and Data, 2) Critical Thinking and Judgment, 3) Ethical and Responsible Use, 4) Human-Centricity, Emotional Intelligence, and Creativity, and 5) Domain Expertise\cite{DEC_AILiteracyFramework_2025}. 
 
 Even before the emergence of AI chatbots, there was concern over the public's limited understanding of AI technologies and opaque algorithms, especially considering their creation of "folk theories" about them \cite{eslami2019CHIopaquealgs,long2020CHIAIlit}. In 2021 Ng's literature survey indicated that researchers advocated for students to understand underlying AI concepts for their careers and to be responsible ethical citizens ~\cite{ng2021ai}. However, in Computer Science courses today, many are diving into using AI without first teaching students how it works. As stated by Miranda, "Simply knowing how to use AI is not enough." ~\cite{miranda2026science}. Our approach seeks to explicitly incorporate the demystification of LLMs, both how they work and how they can influence learning, so students can establish an informed mental model.

\subsection{Use}
 CS instructors have taken a range of approaches to respond to LLM use, including banning it, allowing unrestricted use, or adopting moderated policies in between \cite{ bouvier2025rest, denny2024cacm}. Some instructors redesign assignments or policies to account for AI capabilities, including restricting use or adapting coursework expectations \cite{prather2025}. Others integrate LLM chatbots with guardrails or structured activities, such as guided prompting, code explanation, or evaluation tasks, to support learning rather than replace it \cite{kazemitabaar2024codehelp, kumar2024guiding, long2025acat}.
 
 Research has shown that students use LLMs for a wide range of programming tasks including understanding, debugging, and generating code~\cite{ amoozadeh2024student,  bernstein2025beyond, Utkarsh2025}. However, in~\cite{Utkarsh2025}, Arora et al. reported 37\% of interactions with LLMs involved pasting entire assignment descriptions which may lead to shallow learning. In~\cite{bernstein2025beyond}, students themselves reported this risk by mentioning that LLMs, while helping sometimes, can reduce human power and human thinking ability. We aimed to update our CS2 course to provide experiences for students to use LLMs but with enough boundaries so that students learn to do so effectively and ethically while still being able to meet the course learning outcomes without assistance.

 \subsection{Reflect}

Reflection assignments are a well-known approach for improving student meta-cognition and academic self-regulation ~\cite{zimmerman2002becoming}. In 2011 VanDerGrift et al incorporated a reflection assignment to help students become more meta-cognitive about their own software development processes~\cite{vandegriftreflection2011} and by 2020 many CS educators had explored its use ~\cite{prather2020reflection}. Recently, STEM education researchers are exploring the use of AI as a meta-cognitive partner ~\cite{lowry2025leveraging,prasad2024self}. Padiyath has created a GenAI contract tool to help programming students set goals on AI use, and reflect on their alignment to those goals ~\cite{padiyath2026genaicontracts}. In 2024, Margulieux et al. asked students in an introductory programming course to reflect on their process for completing assignments with AI repeatedly \cite{margulieux2024}. Our approach similarly aims to use reflections to mitigate the risk of students using LLMs mindlessly and encourage them to increase their awareness, critical use, and effectiveness over time. Our goal is to support both productive use of AI and development of student critical thinking. Furthermore, current industry trends indicate that communication skills are becoming even more important.

\subsection{Assess}
Many instructors have adjusted assessment practices by switching to in-class, and even handwritten, tests in order to verify student understanding beyond submitted code. For programming assignments, instructors are now often requiring video, written, or in person oral explanations of code \cite{prather2025}.  For example, the 2024 AP Computer Science Principles exam was redesigned to require students describe program design, algorithm behavior, errors, and abstractions \cite{Lee2023APCSPChange}. Given our emphasis on students meeting learning outcomes unassisted, we adapted our in class proctored assessments to be more aligned with programming assignments and allowed retakes.

Unlike policies that restrict or monitor LLM use, the approach we describe allows the use of LLMs but with assessment design to provide guardrails against over-reliance. While many instructors are exploring a wide variety of uses and integration approaches for LLMs, DURA fills a gap because it also presents complementary approaches. This framework, and especially our emphasis on demystifying LLMs in a programming course, are novel contributions.

\section{Course Description}
We updated a CS2 course at our large research-intensive institution in the southeast United States where the Department of Computer Science is part of the College of Engineering.
The course was taught in Java and covered an introduction to data structures with an object-oriented approach over a 15-week semester and emphasized test-driven development. The course covered the use and implementation of the introductory data structures bags, stacks, queues, lists, and binary trees. It is a prerequisite for required core CS courses Computer Organization and Data Structures and Algorithms as well as several CS elective courses. In the fall of 2025 there were three instructors across four lecture sections teaching 442 students, of whom 53\% were CS majors or first-year General Engineering majors and the others were spread across various majors with 23\% from Data Science. 
It was a flipped course that used an interactive e-book which included instructional videos. During lecture, instructors gave demos, highlighted concepts with Peer Instruction as shown to be effective by Porter et al. \cite{porterPI2014}, and provided time for learning activities. The team of 18 TAs led lab sections where students worked on weekly lab programming assignments or other coursework. Instructors held drop-in office hours and TAs offered centralized in-person office hours weekdays from 12 - 8pm and there was a course Piazza ~\cite{piazza} forum monitored by instructors and TAs. TAs attended a departmental workshop at the beginning of the semester, and the instructional team met weekly for one hour and communicated regularly using a chat system.

\section{Course Update Approaches}
In an effort to prepare students for programming with access to LLMs, we adjusted the course policy to allow students to seek external assistance such as from LLMs. With an eye towards the future, we expected these changes to be the first steps as we adapted to the changing AI landscape. They were manageable initial steps before we were ready to invest efforts to overhaul labs and projects. Instructors can easily adapt our materials and approaches to incorporate the demystify, use, and reflect components into any CS course. However, adjusting assessments to include programming assignment skills is course specific and can be a more time-intensive undertaking, especially when creating question banks to support retakes. We worked on this over the summer with UTA support, as generating question variations was not reliable without oversight. Reusable course materials are available at \url{https://tinyurl.com/3z8kvfnm}.

\subsection{Demystify LLMs}
In the second week of the course, we added a 30-minute introduction to how LLMs work and highlighted the CS education research on how LLM use can affect student learning. The presentation of LLMs explains how LLMs are actually a machine learning model, called a Transformer, which does next token prediction based on the training data. We emphasized that LLMs do not think or problem solve and while it's beneficial that students learn to use them, they should be careful not to short cut their own learning and problem solving skill development.

This lesson was adapted from another introductory course where students spend more time learning details about how LLMs work \cite{donekal2026demystify} and includes information about CS education research which references articles covering the impact of LLMs on student ability to write, trace, and read code as well as the distinctions between higher performing students and those who may not have a solid base \cite{vadaparty2024cs1,margulieux2024,padiyath2024insights,prather2024widening}. An example of the risks for a CS2 student using an LLM for programming generated by ChatGPT, including shallow understanding and inhibited debugging skills, was also included. This LLM demystification lesson can be used in a variety of courses.

\subsection{Responsible LLM use}
 Students were also provided with some LLM suggested use guidance both during the lesson and on a course page. During the lesson, student responsibilities to work through course assignments, document use of LLMs, and meet learning objectives unassisted were highlighted. We suggested strategies for LLM use such as trying to solve problems on your own first using tracing, diagrams, and the debugger. We encouraged students to seek guidance from course staff and that when they do use an LLM to be clear and specific, go step by step, and verify responses.

 A course page had reminders to caution students about short-cutting learning as well as example prompts to encourage students to use LLMs as a learning aide. We also discussed and provided sparse and detailed prompt examples of how a students could instruct an LLM to be an AI study assistant and prioritize learning over giving answers.

 Throughout the course, students are expected to provide an attribution statement when using LLMs. The expectation is that each code file should include one of the following:

\begin{itemize}
    \item I have not used any assistance for the assignment.
    \item During the preparation of this assignment, [name of the student] used [name of the tool] in [name the parts of the assignment where it was used] to [specify the purpose]. After using this tool, I/we reviewed and edited the content as needed to ensure its accuracy and take full responsibility for the content in relation to grading. I understand that I am responsible for being able to complete this work without the use of assistance.
\end{itemize}

Note that the inclusion of the last sentence is important for this course design, the goal was for students to still be able to meet learning objectives without assistance.

\subsection{Student Reflection Assignments}

We added problem solving reflections, where students were asked to select a recent programming assignment and describe their process in their own words, in weeks 4, 8, and 13. In week 4 students were also asked to guide their former self on how to approach the assignment. In week 8, students were also asked to reflect on whether they followed their previous advice.  In week 13, after the final individual project was due, students were also asked give advice to future students.
 
 The TAs graded these reflections which stimulated valuable discussions about student perspectives, help-seeking, and LLM use, during weekly instructional team meetings. After the Week 4 Problem Solving Reflections, some TAs were surprised to read that students were intimidated by them. Although we have always explained this to TAs, reading it in students' own words made an impression on the TAs. To better understand student perceptions of help-seeking strategies, our team analyzed the Week 13 Problem Solving Reflections which covered the most challenging individual programming projects. 
As shown in Table \ref{tab:wk13}, the assignment included two sets of questions designed to capture students’ reflections. We created a summary of the advice for use in lecture that following semester.

\begin{table}[h]
\centering
\small
\caption{Week 13 Problem Solving Reflection Assignment}
\begin{tabular}{p{26em}}
\hline
\textbf{Assignment} \\ \hline

\textbf{1. Project Reflection (200--250 words)} \\ 
For either Project 3 or Project 4, tell the story of how you came up with your solution. Be sure to include: 
\begin{itemize}
    \item the timing of when you began the process and what your first steps were
    \item details on how, when, and why you sought assistance or did not seek assistance (e.g., Piazza, office hours, LLMs)
    \item if you used a large language model (LLM), describe your strategy for using it; if not, explicitly indicate that
    \item an explanation of any hurdles you encountered (e.g., moments that required rethinking or repeated attempts)
\end{itemize} \\ \hline

\textbf{2. Advice to Future CS2114 Students (100--150 words)} \\ 
Provide advice addressing the following:
\begin{itemize}
    \item how you recommend students approach course assignments, problem solving, and test preparation
    \item how, when, and why you recommend students seek assistance
\end{itemize} \\ \hline

\end{tabular}
\label{tab:wk13}
\end{table}

\subsection{Assessment Updates}
We reallocated 20\% of the course grade from programming assignments (45\%->25\%) to proctored assessments (35\% -> 55\%) to emphasize independent comprehension. To offset this shift to high-stakes testing and encourage student learning, we allowed an in-class proctored retake of each test during the semester ~\cite{baralquizretake2021}. Recent versions of the course had 4 in-class proctored quizzes, so the course was rearranged to use the four class days for 2 tests each with a retake day one week later.

The questions related to programming assignments often included method javadocs or test methods closely aligned with assignments. Because the tests were multiple choice they did often require careful inspection of code by students. The intention was that students who engaged with the code when working on the assignments would be well prepared for the tests.

Tests were created from banks of questions. Each week students had sample test questions to optionally practice. After the first attempt at a test, the students could see their test score, but needed to come to office hours to look through their tests with a TA or Instructor. It was a good learning opportunity for students and also seemed to help them build relationships with the course staff. This was a strong motivator to get students to attend office hours and possibly realize it was useful to seek in-person assistance.

\subsection{Instructor Impressions}
Two of the three instructors during Fall 2025 are co-authors of this report. We enjoyed the changes to the course, it was refreshing to be able to openly discuss LLM use with students and there was no longer a sense that we were policing them. It also opened the door to more conversations about the hows and whys of the course design, thus creating a more transparent teaching environment ~\cite{10.1145/3770761.3777041}. Because of this foundation, we were also well situated to discuss AI advancements and current events with the students. Some TAs expressed concern that the students were using LLMs and over-reliant on them. After the first test some students realized they needed to take their assignments more seriously and we were able to have open dialogue with them about this, especially when they came to review their tests. 

In the future, we would like to have assessments earlier and more often.
In general, we were encouraged that some students seemed to prefer talking with us and attending office hours, even though LLMs were allowed. As AI coding agents become more advanced, we would like to provide students access and demonstrations of coding agents in an environment that encourages student effective use of LLMs. This involves re-imagining CS2-level programming assignments and use of lab time to bolster student engagement with peers and TAs.

\section{Evidence}

We observed a change in student behaviors and perceptions as explained in more detail below, but we cannot attribute causality due to lack of a control group. We have IRB approval for the use of student data in this course.

\subsection{Course Help-Seeking History}
\begin{figure}[h]
  \centering
  \includegraphics[width= 0.85\linewidth]{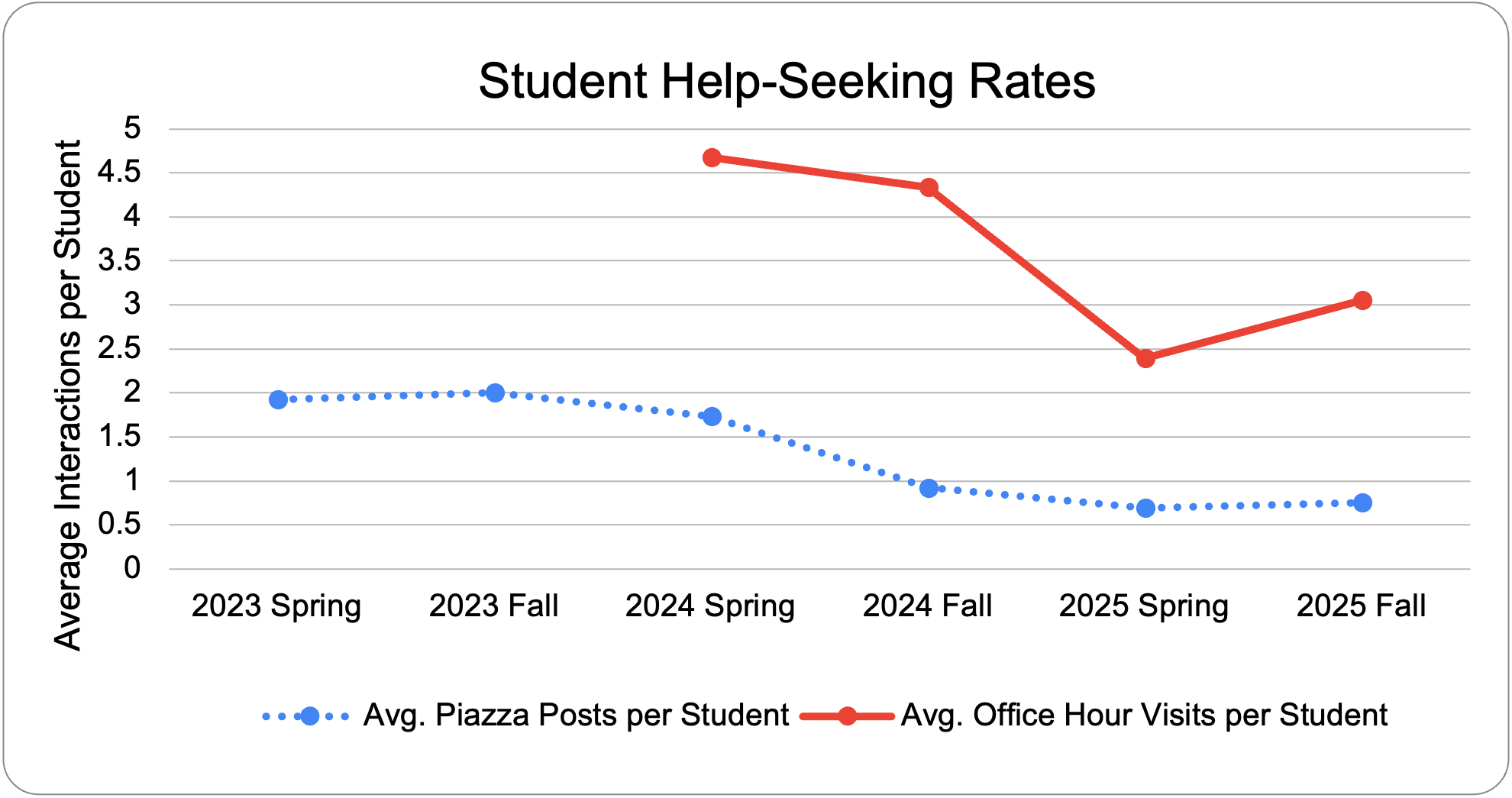}
  \caption{Student interaction rate by average piazza post rate per student and average office hours visit per student}
  \Description{Piazza rate drops from above 2.5 to below 1 from 2021 Fall to 2025 Fall and Office hours rate drops from above 4.5 in 2024 Spring to 3 in 2025 Fall).}
  \label{fig:piazza_officehours}
\end{figure}

As in other CS courses ~\cite{amoozadeh2024student, hou2024effects}, our team had observed the reduced number of students attending office hours and posting on the Piazza course forum. The decrease in student engagement in recent years in this CS2 course is shown in Table \ref{tab:course_activity_by_semester}.
Although use of LLMs was not sanctioned until 2025 Fall, notice in Figure \ref{fig:piazza_officehours} that there was steep decline in Piazza posting in 2024 Fall and office hours usage in 2025 Spring and then a rise in Fall 2025 when LLMs and retakes were allowed. To better understand student perceptions of help-seeking, our team analyzed the student responses for the Week 13 Problem Solving Reflections assignment.

\begin{table}[h]
\centering
\small
\caption{Student help-seeking rates by semester: enrollment, Piazza activity, and office hours interactions}
\label{tab:course_activity_by_semester}
\begin{tabular}{lrrrrl}
\toprule
\textbf{Semester} & \textbf{Enrollment} & \textbf{Piazza posts} & \textbf{Office hours}  \\
\midrule
2023 Spring & 433 & 834  & N/A  \\
2023 Fall & 619 & 1239 &  N/A  \\
2024 Spring & 571 & 990  &  2672  \\
2024 Fall & 597 & 551  &  2588  \\
2025 Spring & 480 & 334  &  1151  \\
2025 Fall & 442 & 334  &  1350  \\
\bottomrule
\end{tabular}

\end{table}

\subsection{Student Reflections}
Of the 442 students enrolled in the course, 198 students provided informed consent and submitted valid responses for the Week 13 Problem Solving Reflection assignment, with prompts shown in Table \ref{tab:wk13}.

A team of five co-authors, two of whom were instructors for the course, analyzed the anonymized responses using an iterative qualitative coding approach that combined inductive and deductive coding consistent with established practices \cite{saldana2016qualitative, cohen2024factors}. 
The final codebook consisted of 32 codes related to students’ help-seeking behaviors as shown in Table \ref{tab:help_seeking_codes_single}, and 35 codes related to students’ recommendations for future help-seeking as shown in Table \ref{tab:help_seeking_advice_codes_single}.

\begin{table} [b]
    \centering
    \small
        \caption{Qualitative Codes for Student Help Seeking Behavior}
    \begin{tabular}{p{26em}}
        \toprule
        \textbf{Theme / Code} \\ \midrule

        \textbf{Supplemental:} Classmates/Friends, Online Resources \\

        \textbf{Self-Driven:} Debuggers, Code Tracing, Test Cases \\

        \textbf{LLM for Understanding:}
        \textit{Instructions} (advice/check direction, break down, clarify);
        \textit{Concept Understanding} (clarify concepts, tutor);
        \textit{Technical Assistance} (IT assistance, Java syntax) \\

        \textbf{LLM to Assist with Code Writing:}
        Debugging, Testing, Comments/Documentation/Javadocs, Refactoring Suggestions \\

        \textbf{LLM for Code Writing:}
        LLM for Code Writing \\

        \textbf{Formal Resources:}
        \textit{Course Materials} (assignment instructions, course content, videos);
        \textit{Instructional Team} (TA, Professor) \\

        \textbf{Modality:}
        \textit{Piazza} (read-only, generic);
        \textit{General} (labs, lecture, office hours) \\

        \textbf{Selective Help Seeking:}
        No LLM, No Office Hours, No Piazza, None (choice or lack of time) \\

        \textbf{Generic:}
        Help Seeking (Generic) \\
        \bottomrule
    \end{tabular}

    \label{tab:help_seeking_codes_single}
\end{table}

Students reported using formal course resources such as office hours, Piazza, TAs, and course materials. Some students did not seek or use any assistance and most stated that no LLM was used for the assignment described.
\begin{quotation}
S40: \textit{ I did not use an LLM as it did not help fix the coverage and just took me down a rabbit hole that never even helped.}
\end{quotation}
\begin{quotation}  
S210: \textit{I did not use assistance from LLMs, but rather went to office hours since it was much more helpful to talk to someone face to face and get a well-descriptive explanation.}
\end{quotation}

Many students did seek assistance from LLMs and students described a variety of approaches for understanding, code writing assistance, and code writing that are similar to those found in prior work ~\cite{hou2024effects, amoozadeh2024student,bernstein2025beyond, Utkarsh2025}. Students commonly used multiple resources and approaches. How, when, and why students did use LLMs varied, their usage can be categorized into the themes: used to understand, used to assist code writing, and used to write code. 

\begin{quotation}
S12: \textit{I used an LLM as a support tool, but not for writing code. My strategy was to ask it to rephrase confusing parts of the spec, help me walk through example queue states, and think through potential edge cases.}
\end{quotation}
\begin{quotation}
S226:\textit{ I also used an LLM as a debugging partner and have it explain concepts to me. An example was asking questions like "why wasn't this comparator working" or "has this test fully covered the code?" 
}
\end{quotation}
Only a few reflections described using LLMs to write the code.
\begin{quotation}
S232:\textit{ I used to be completely against the use of AI in classwork, but in this project I saw the value of consulting an LLM for outlines and explaining methods I didn’t understand how to implement...I did change the way I would’ve implemented some methods based on Claude suggestions, which helped me think more efficiently in certain aspects, but in others I stuck with my own implementation and did not agree with the AI’s thoughts.}
\end{quotation}
\begin{quotation}  
S10:\textit{ I did use LLMs to help me figure out some more complex parts of the constructors/method as well. I would insert the prompt that I was given on canvas, along with my personal thought process. It would result in the LLM giving me a code result with an explanation that helped me understand why it was writing that code and how it worked.
}
\end{quotation}

\begin{table}[h]
    \centering
    \small
        \caption{Qualitative Codes for Student Help Seeking Advice}
    \begin{tabular}{p{26em}}
        \toprule
        \textbf{Theme / Code} \\ \midrule

        \textbf{Supplemental:}
        Classmates/Friends, Online Resources \\

        \textbf{Self-Driven:}
        Debuggers, Code Tracing \\

        \textbf{LLM for Understanding:}
        LLM not for code writing;
        \textit{Instructions} (break down, clarify);
        \textit{Concept Understanding} (clarify concepts, tutor) \\

        \textbf{LLM to Assist with Code Writing:}
        Debugging (clarifying/resolving errors), LLM as assistant, Testing \\

        \textbf{Why:}
        Understand process/concept, Getting unstuck, Testing-related queries, IT technical assistance \\

        \textbf{Formal Resources:}
        \textit{Course Materials} (assignment instructions, course content);
        \textit{Instructional Team} (TA, Professor) \\

        \textbf{Modality:}
        \textit{Piazza} (read-only, generic);
        \textit{General} (labs, lecture, office hours) \\

        \textbf{Selective Help Seeking:}
        No LLM, No Office Hours, No Piazza \\

        \textbf{When:}
        After reviewing course materials, Anytime struggling, Try on own first,
        When stuck for a duration, Start early to allow time to seek help \\
        \bottomrule
    \end{tabular}

    \label{tab:help_seeking_advice_codes_single}
\end{table}

Students mostly recommended future students seek help via formal help modalities and the course staff but also supplemental sources such as friends and online resources. Similar to Penney's findings~\cite{penney2025understanding}, students predominantly recommended that future students use office hours and they commented on how helpful the TAs were. Piazza was also recommended and students suggested using the course materials for help.
\begin{quotation}  
S194:\textit{ I highly recommend going to office hours and talking to TAs whenever something feels confusing or even in general because they're only there to help. Even when you think you understand everything, going to office hours can is so helpful because it teaches you so much. 
}
\end{quotation}
\begin{quotation}  
S123:\textit{ 
If a student is stuck, I would 
recommend them to ask their TAs or go to office hours before trying LLMs, because they really want to help you, and they might explain the problem better than LLMs.}
\end{quotation}

How, when, and why students recommended getting assistance varied. There was a noticeable trend to recommend starting assignments early in order to be able to get help and not be overwhelmed. Students also explicitly mentioned that students should try to solve problems on their own first or look through course materials. Students described the main reason why a future student should seek assistance is to understand the material.

\begin{quotation}
    S121: \textit{
    Having the knowledge of how a solution works should be more important than just getting the answer correct or getting the assignment done. }
\end{quotation}
\begin{quotation}  
    S192:\textit { To understand how code works is the most important, and if you push off the work by using LLMs (or other outside assistance) you will be unprepared for the tests.}
\end{quotation}

Students recommend using LLMs to assist with code writing such as for test cases or documentation. But some students cautioned the use of LLMS.
\begin{quotation}
S36:\textit { Use Piazza for quick clarifications, office hours for design questions, and LLMs to rephrase specs or suggest test cases—then make sure you still understand and test every line you submit.}
\end{quotation}
\begin{quotation}
S43:\textit{ My biggest advice to new students is to avoid using AI to do large parts of the work as it makes it harder to get a real feel for doing the code yourself
}
\end{quotation}

 Student help-seeking was not simply AI and non-AI behaviors. There is a whole gamut of possibilities of how, when, and why students do and do not use LLMs. Increased instruction on incorporating LLMs into workflows may benefit some students.

\subsection{Student Perceptions of Caring}
We also compared student perceptions of the motivational climate for this CS2 course to the previous fall semester using the MUSIC Inventory as shown in Table \ref{tab:MUSIC}. This questionnaire is based on The MUSIC Model of Motivation and measures student perceptions of course eMpowerment/autonomy, Usefulness, Success, Interest, and Caring using a Likert Scale from 1(Strongly Disagree) - 6 (Strongly Agree) \cite{JonesMUSICQuestionnaires}. Student MUSIC perceptions were quite consistent but students reported significantly higher perception of instructor Caring (p < .001) during the semester we made these DURA course updates. The 4 questions for instructor Caring on the MUSIC Inventory are:
\begin{itemize}
    \item The instructor is willing to assist me if I need help in the course.
    \item The instructor cares about how well I do in this course.
    \item The instructor is respectful of me.
    \item The instructor is friendly.
\end{itemize}

Although multiple factors may have an impact, allowing assessment retakes along with the shift away from reporting students for honor code violations to guiding and encouraging responsible use of assistance may have been perceived as more caring.

\begin{table}[h]
\centering
\caption{Comparison of MUSIC Survey Results by Term}
\resizebox{\columnwidth}{!}{%
\begin{tabular}{lccccccccc}
\hline
Term & Weeks & $n$ & $N$ & \% & M & U & S & I & C \\
\hline
2025 Fall & 15 & 235 & 442 & 53.2 & 4.6 & 5.0 & 4.9 & 4.4 & \textbf{5.3*} \\
2024 Fall & 15 & 329 & 597 & 55.1 & 4.6 & 5.0 & 5.0 & 4.2 & \textbf{4.7} \\
\hline
\end{tabular}
}
\textit{Note.} $n$ = number of respondents; $N$ = total enrollment; \% = response rate. * p < .001
\label{tab:MUSIC}
\end{table}

\section{Conclusion}
In this experience report we detailed course updates to promote effective and ethical use of LLMs at scale. Our materials to demystify, use, and reflect on LLMs are available and can be used in CS courses at various levels and adapted to reflect instructor policies and goals. Our use of in-person proctored tests that addressed programming assignment skills provided a guardrail against over-reliance of LLMs and requires course-specific implementation. Likewise supporting test review and retakes requires support by the instructional team. Allowing both LLM use and test retakes helped create a more supportive and open course culture.

\begin{acks}
We used ChatGPT to explore related work, add the arcs and circle to the DURA diagram, and provide feedback on the report contents.
\end{acks}


\bibliographystyle{ACM-Reference-Format}
\bibliography{LLMReady}

\end{document}